\documentclass{pasj01}
\usepackage{color}
\draft

\Received{$\langle$reception date$\rangle$}
\Accepted{$\langle$acception date$\rangle$}
\Published{$\langle$publication date$\rangle$}
\newcommand{\thirteencoh}{\mbox{$^{13}$CO($J$ = 2--1)}} 
\newcommand{\twelvecol}{\mbox{$^{12}$CO($J$ = 1--0)}} 
\newcommand{\thirteencol}{\mbox{$^{13}$CO($J$ = 1--0)}}

\newcommand {\msun}{\mbox{$M_\odot$}}
\newcommand {\kms}{\mbox{km~s$^{-1}$}}
\newcommand {\kkms}{\mbox{K~km~s$^{-1}$}}

\newcommand {\nhtwo}{\mbox{$N_\mathrm{H_2}$}}

\newcommand {\tastar}{\mbox{$T_\mathrm{a}^{*}$}}

\begin{document}

\title{A kinematic analysis of the CO clouds toward a reflection nebula NGC~2023 observed with the Nobeyama~45~m telescope; Further evidence for a cloud-cloud collision in the Orion region}
%\author{$^1$Rin Yamada, $^1$Rei Enokiya, $^1$$^,$$^2$Shinji Fujita, $^1$$^,$$^2$Hidetoshi Sano, $^1$$^,$$^2$Daichi Tsutsumi, $^1$Mikito Kohno, $^1$$^,$$^3$Katsuhiro Hayashi ,$^1$Kengo Tachihara, and $^1$$^,$$^4$Yasuo~Fukui}%

\author{Rin \textsc{Yamada}\altaffilmark{1}, Rei \textsc{Enokiya}\altaffilmark{1}, Hidetoshi \textsc{Sano}\altaffilmark{1, 2}, Shinji \textsc{Fujita}\altaffilmark{1, 3}, Mikito \textsc{Kohno}\altaffilmark{1, 4},Daichi \textsc{Tsutsumi}\altaffilmark{1}, Atsushi \textsc{Nishimura}\altaffilmark{3}, Kengo \textsc{Tachihara}\altaffilmark{1}, and Yasuo \textsc{Fukui}\altaffilmark{1, 5}}

\altaffiltext{1}{Department of Physics, Nagoya University, Furo-cho, Chikusa-ku, Nagoya 464-8601, Japan}

\altaffiltext{2}{National Astronomical Observatory of Japan, National Institutes of Natural Sciences,
2-21-1 Osawa, Mitaka, Tokyo 181-8588, Japan}

\altaffiltext{3}{Department of Physical Science, Graduate School of Science, Osaka Prefecture University, 1-1 Gakuen-cho, Naka-ku, Sakai, Osaka 599-8531, Japan}

\altaffiltext{4}{Astronomy Section, Nagoya City Science Museum, 2-17-1 Sakae, Naka-ku, Nagoya, Aichi 460-0008, Japan}

\altaffiltext{5}{Institute for Advanced Research, Nagoya University, Furo-cho, Chikusa-ku, Nagoya 464-8601, Japan}

\email{yamada@a.phys.nagoya-u.ac.jp}

\KeyWords{ISM : clouds${}$ --- ISM : kinematics and dynamics ${}$ --- ISM : Molecules --- stars : formation${}$}

\maketitle

\begin{abstract}
We have analyzed new CO($J$ = 1--0) data in the region of a reflection nebula NGC~2023 with a particular focus on the detailed kinematical properties of the molecular gas. The results show that there are two velocity components which indicate signatures of dynamical interaction revealed at a high resolution of 19$\arcsec$  (= 0.04 pc). Based on the results we propose a hypothesis that two clouds collided with each other and triggered the formation of the B1.5 star HD~37903 in addition to 20 lower mass stars in two small clusters with a size of 2 pc. Although the previous study favored a scheme of triggering by the H{\sc ii} region (e.g., \cite{2009A&A...507.1485M}), the present results show that the effect of the H{\sc ii} region is limited only to the surface of the molecular cloud, and does not contribute to the gas compression and star formation. The present results lend support for the dominant role of cloud-cloud collision in forming high mass stars in addition to $\sim$20 lower mass stars, which are also likely formed by the collision. The present case suggests all the high mass stars in the Orion region are formed by cloud-cloud collision.
\end{abstract}

\section{Introduction}
The H{\sc ii} regions ionized by O or early B stars can compress the neutral gas by high pressure, which was first suggested by \citet{1954BAN....12..177O}, and the role of H{\sc ii} regions in gas compression was studied into detail by \citet{1955IAUS....2..115K}. The gas compression driven by H{\sc ii} regions was proposed by \citet{1977ApJ...214..725E} as a mechanism which triggers sequential formation of OB sub-groups. A number of works were made along the line including the H{\sc ii} driven bubble like RCW~120 \citep{2005A&A...433..565D}. Before that, \citet{1946MNRAS.106..159O} made another suggestion that the interstellar clouds are colliding often in the interstellar space based on optical absorption lines of a few discrete clouds moving at 10--15 \kms, and the collision between clouds or cloud-cloud collision (CCC) may be an important process. This possibility was however not much followed until recently. Among the works of CCC on a various scale sizes (\cite{2011ApJ...738...46T}, \yearcite{2015ApJ...806....7T}, \yearcite{2017ApJ...835..142T}; \cite{2014ApJ...780...36F}, \yearcite{2015ApJ...807L...4F}), \citet{1992PASJ...44..203H} and \citet{2013ApJ...774L..31I} presented theoretical works on CCC and showed that CCC is a viable mechanism to compress gas to trigger the formation of high-mass stars. \citet{2018ApJ...859..166F} presented a methodology to identify CCC, where the following three features are used as observational signatures of CCC; i. two independent clouds having supersonic velocity separation are associated with high mass stars, ii. they show complementary distribution in space, and iii. they are linked by bridge features, which often show a V-shape in a position-velocity diagram.
Until now more than 50 cases of CCC are found in the literature (see a compilation by \cite{2019PASJ..tmp..127E}).

The Orion region is a unique site of active star formation within 400 pc of the sun, and serves as a precious site for studying star formation in detail, in particular, high-mass star formation. The most outstanding high-mass star cluster the Orion Nebula Cluster (ONC) in M42 is the best studied cluster in the Galaxy and the other high mass stars are distributed over 100 pc along the Galactic plane, which include M43, NGC~2024, NGC~2068, and NGC~2071 (see for a review of star formation \cite{1989ARA&A..27...41G}). According to the recent results, high-mass star formation triggered by CCC is shown in M42/M43 (\cite{2018ApJ...859..166F}), NGC~2068/NGC~2071 (\cite{2020PASJ..tmp..163F}), and NGC~2024 (\cite{2019arXiv191211607E}).

NGC~2023 is a reflection nebular in L1630, which is illuminated by an early B star HD~37903, and is separated by about 4 pc from the nearby H{\sc ii} region NGC~2024. 
Star formation in NGC~2023 has been extensively studied by a number of authors. \citet{1990ApJ...356L..55D} made JHK observations and identified 16 protostars at 1\farcs3 resolution with 1.2 m telescope of Kitt Peak National Observatory. About a half of them are subject to the interstellar reddening, while the other half are subject to reddening due to disks or envelopes. \citet{1991ApJ...368..432L} identified a cluster consisting of 21 members by a survey at 2.2 mm with the same instrument. A mm CS($J$ = 2--1) survey at 1\farcs8 resolution revealed five dense molecular cores LBS~34, 35, 36, 39, and 42 \citet{1991ApJ...371..171L}. Observations with the SEST 15 m telescope and IRAM 30 m telescope of the 1.3 mm continuum radiation at resolutions of $23\arcsec$ and $12\arcsec$, respectively, two sub-mm sources LBS36 SM1 and SM2 were discovered \citep{1996A&A...312..569L}. \citet{1999ApJ...519..236S} discovered molecular outflow whose velocity span is more than 200 \kms ~with a dynamical time scale of less than 3000 yr in observations of CO($J$ = 2--1) and CO($J$ = 3--2) emission at resolutions of 21\farcm5 and 14\arcmin, respectively, with the James Clerk Maxwell Telescope (JCMT), and called it NGC~2023 MM1. Another sub-mm source NGC~2023 MM2 was found to be associated with LBS36 SM2 by the same author. \citet{2006ApJ...639..259J} made a survey at 850 $\mu$m at $3\arcsec$ resolution and identified 23 sources in the region of NGC~2023 and the Horse Head Nebula. \citet{2009A&A...507.1485M} identified 73 YSOs based on the data of {\it{Spitzer}} IRAC and MIPS and the 850 $\mu$m and 450 $\mu$m results of JCMT Submillimetre Common-User Bolometer Array. They made SED fitting and identified five sources to be Class 0. These observations also revealed a class I/0 associated with a sub-mm source NGC~2023 MM3 \citep{2000ApJ...543..245W} and that NGC2024 MM4 shows properties of a cold protostellar core. The authors also identified 10 class I/II sources, and 58 class II/III and class III sources. NGC~2023 is faced to an H{\sc ii} region IC~434 and the compression by the H{\sc ii} region was discussed as a trigger of star formation by \citet{2009A&A...507.1485M}.

The molecular clouds in NGC~2024 was studied by a number of authors (e.g., \cite{2008hsf1.book..662M}). Recently, \citet{2013MNRAS.431.1296R} used the FCRAO 14m telescope to map the CO($J$ = 1--0) emission in NGC~2023 with a $45\arcsec$ resolution. \citet{2015ApJS..216...18N} made an extensive large-scale study of the CO($J$ = 1--0) and CO($J$ = 2--1) emissions of $^{12}$CO, $^{13}$CO, and C$^{18}$O obtained with the Osaka Prefecture University 1.85 m telescope and NANTEN, and revealed the detailed excitation states of CO. Most recently, the CO observations by \citet{2019arXiv191211607E} were interpreted to show CCC as a trigger of O star formation. It is therefore interesting to explore the possible mechanism of triggering in NGC~2023 in order to have a comprehensive understanding of the L1630 region. 

The aim of the present work is to make a detailed analysis of the CO cloud toward NGC~2023 and test what is the trigger of star formation. \citet{2015ApJS..216...18N} and \citet{2018ApJ...859..166F} showed that the Orion region have multiple velocity components in a large scale. \citet{2019arXiv191211607E} noted that the small velocity difference of $\sim$2 $\kms$ among these components requires a velocity resolution higher than 1 $\kms$. The NGC~2023 molecular cloud has not been observed at spatial high resolution better than $30\arcsec$ and with a high velocity resolution better than 1 $\kms$. The present study aims at revealing detailed kinematics of the molecular gas with the NRO 45 m telescope at $19\arcsec$ resolution and 0.3 $\kms$ velocity resolution. The directions in the sky are given in the Galactic coordinates in the present paper. This paper is organized as follows. Section \ref{sec:obs} describes the CO observations. Section \ref{sec:results} presents the results, and Section \ref{sec:discussion} discussion on star formation by triggering based on detailed study of cloud kinematics. Section \ref{sec:conclusions} gives conclusions.

\section{Observations}\label{sec:obs}

\subsection{Observations by the Nobeyama 45-m telescope}
We performed four molecular line observations including $\twelvecol$ and $\thirteencol$ line emission with 45-m telescope at the Nobeyama radio observatory in January 14, 2017 to 15, 2017 together with NGC~2024 observations (\cite{2019arXiv191211607E}). The half-power beam width is 14$\arcsec$ at 115~GHz, and 15$\arcsec$ at 100~GHz. We scanned $22\arcmin \times 26\arcmin$ area with OTF mapping mode. The map center was ({\it{l, b}}) = (206\fdg86, $-16\fdg53$). To release scanning effect, we scanned $l$ direction and $b$ direction and adopted the RMS weighted average. The frontend was four- beam, dual-polarization, and two sideband (2SB) receiver named FOur-beam REceiver System on the 45-m Telescope 
\citet{2016SPIE.9914E..1ZM}. The typical system noise temperatures including atmosphere were 180--270 K at 110 GHz and 150--250 K at 115 GHz in the stellar direction. The backend was SAM45, which is a FX type digital spectrometer \citet{kuno}. SAM45 has 4096 channels with a bandwidth and resolution of 250~MHz and 61.04~kHz. These frequencies correspond to 650 $\kms$ and 0.15 $\kms$, respectively, at 115 GHz. We checked the pointing accuracy before the observation by observing Ori-KL ($\alpha_\mathrm{J2000}$, $\delta_\mathrm{J2000}$) $\sim$ ($05^\mathrm{h}35^\mathrm{m}14$\farcs$5$, $-5^{\degree}22\arcmin30\farcs4$) with the 40 GHz HEMT receiver named H40. The pointing accuracy was better than $3\arcsec$. We used chopper wheel method to gain antenna temperature $\tastar$ \citep{1981ApJ...250..341K}. We observed W3 as a standard source to fix the variation between the four beams in January 5, 2017 to January 7, 2017. The variation between the beams is lower than 20\%. The main beam efficiency ($\eta_\mathrm{mb}$) was 0.44 at 110 GHz and 0.43 at 115 GHz. We carried out two channel binning and convolution, and obtained the final data whose special and velocity resolution were $19\arcsec$ and 0.33 $\kms$. The typical rms noise level was 0.88 K at 110~GHz and 0.37 K at 115~GHz.

\subsection{The other datasets}
In the present paper, we used WISE archive data of 3.5 $\mu$m, 4.5 $\mu$m, and 22 $\mu$m. The special resolution of WISE archive data are $6.4\arcsec$ (3.5 $\mu$m), $6.5\arcsec$ (4.5 $\mu$m), and $12\arcsec$ (22 $\mu$m).

\section{Results}\label{sec:results}
\subsection{The distribution of CO gas}

\begin{figure*}[]
\begin{center}
    \includegraphics[width=\linewidth]{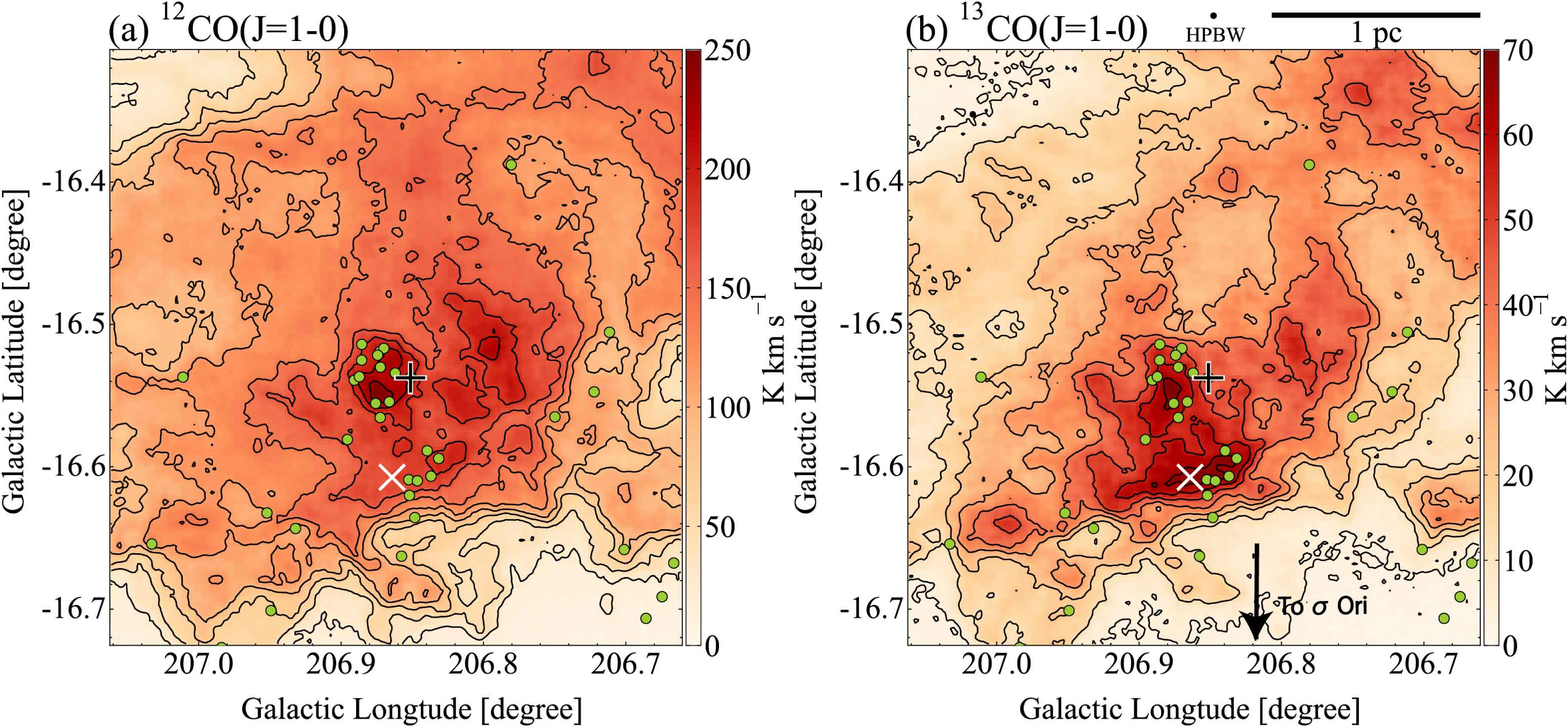}
\end{center}
  \caption{(a) Integrated intensity map of $\twelvecol$ obtained with NRO-45m. The integration velocity range is from 7.9 $\kms$ to 13.9 $\kms$. The lowest contour and contour intervals are 23 $\kkms$ ($\sim$15$\sigma$). (b) Integrated intensity map of $\thirteencol$. The integration velocity range is the same as (a). The lowest contour and contour intervals are 2.9 $\kkms$ ($\sim$5$\sigma$) and 8.7 $\kkms$ ($\sim$15$\sigma$), respectively. The black cross, white cross and green dots represent the positions of HD~37903, NGC~2023 MM1, and YSOs detected by \citet{2009A&A...507.1485M}. }
  \label{integ_int}
\end{figure*}

Figures \ref{integ_int} shows the total intensity distributions of \twelvecol ~and \thirteencol . In Figure \ref{integ_int}a \twelvecol ~distribution has a peak at $(l$, $b)$ = $(206\fdg85$, $-16\fdg55)$, where the exciting star HD~37903 of the reflection nebula and two dust condensations MM3/MM4 are located (\cite{2000ApJ...543..245W, 2009A&A...507.1485M}). The $\thirteencol$ cloud has two peaks at the $\twelvecol$ peak and another peak at $(l$, $b)$ = ($206\fdg83$, $-16\fdg6$), where MM1/MM2 and outflow are located. The $\thirteencol$ cloud is elongated nearly along the Galactic plane at 30 $\kkms$ in Figure \ref{integ_int}b, while it is tilted by 20 degrees to the Galactic plane. The 20 young stars shown by green points are distributed toward the two $\thirteencol$ peaks. We here define the $\thirteencol$ peaks as $\thirteencol$ peak1 toward $(l$, $b)$ = $(206\fdg87$, $-16\fdg55)$ and $\thirteencol$ peak~2 toward $(l$, $b)$ = ($206\fdg85$, $-16\fdg60$). The both CO distributions show sharp intensity decrease toward the south. In $\twelvecol$, $\thirteencol$ peak~2 corresponding to the white cross is not seen probably because of self-absorption. We use the $\thirteencol$ emission in the following, which is not subject to self-absorption.

\begin{figure*}[]
  \begin{center}
    \FigureFile(180mm,90mm){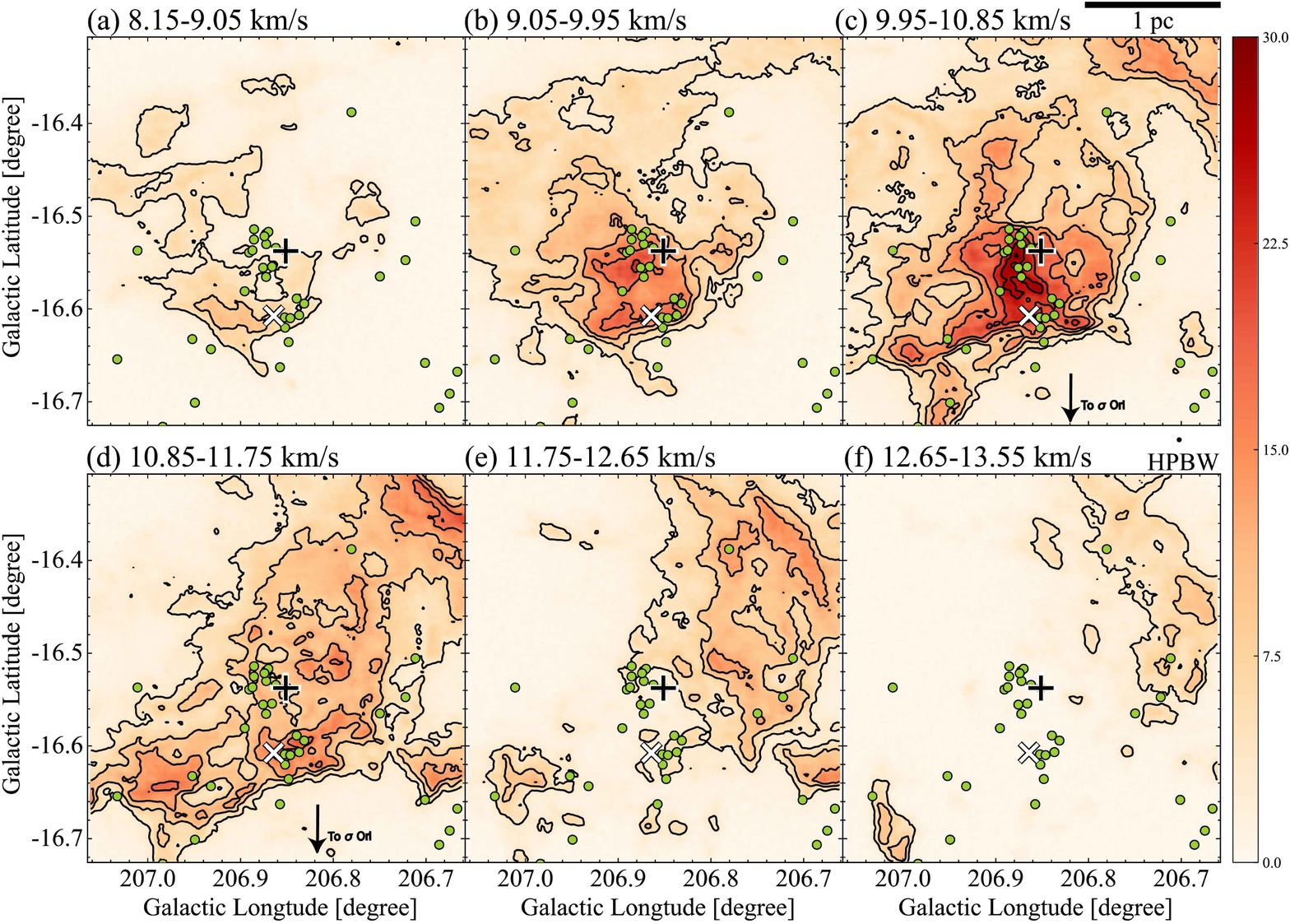}
 \end{center}
  \caption{Velocity channel maps of $\thirteencol$ toward NGC~2023. Each panel shows a $\thirteencol$ intensity map integrated over the velocity range from 8.15 to 13.55 \kms ~every 1.2 \kms . The black cross,  white cross and green dots represent the positions of HD~37903, NGC~2023 MM1 and YSOs detected by \citet{2009A&A...507.1485M}, respectively. The lowest contour and contour intervals are 3.9 \kkms ($\sim$15$\sigma$). }
  \label{chmap}
  \end{figure*}

Figure \ref{chmap} shows the velocity channel distribution of the $\thirteencol$ emission every 0.9 $\kms$ in a velocity range from 8 $\kms$ to 14 $\kms$. The peak~1 is found mainly in panel c, coinciding with the YSO distribution. We see a velocity gradient which show the eastern side is blue-shifted and the western side red-shifted. In panels c and d we see the elongation of the cloud at 20 degrees to the Galactic plane.This coincides with the velocity range of the CO component associated with the H{\sc ii} region IC~434 (\cite{2019arXiv191211607E}).

\begin{figure*}
  \begin{center}
    \includegraphics[width=\linewidth]{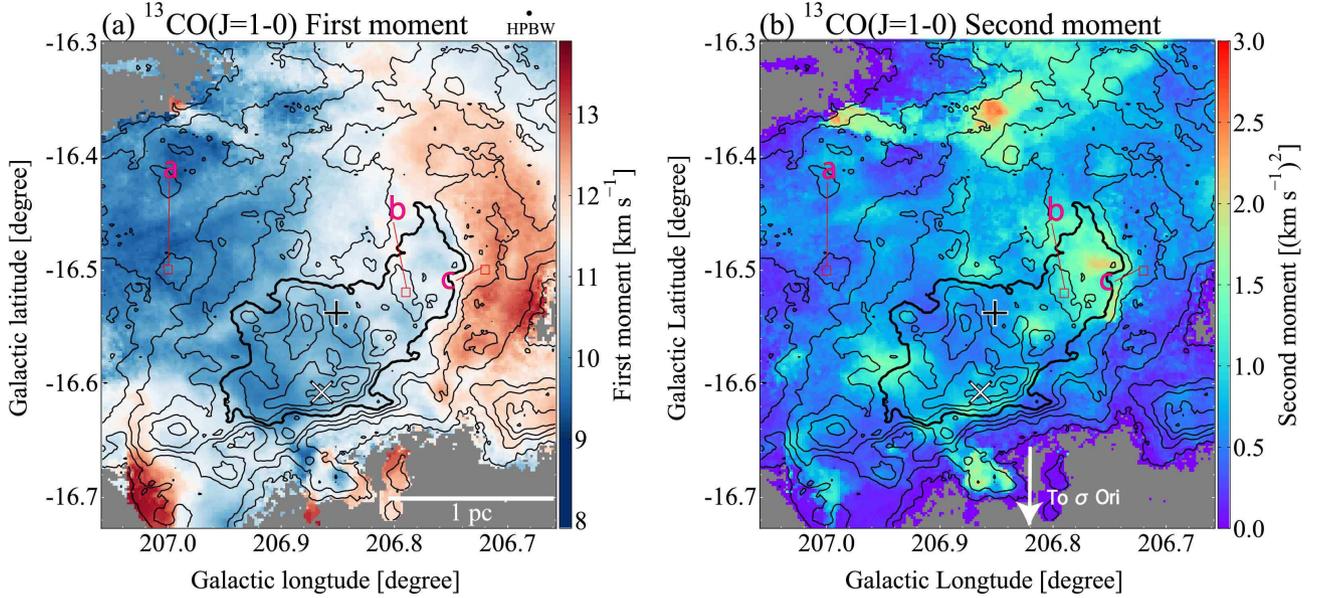}
  \end{center}
  \caption{First moment map (a) and second moment map (b) of $\thirteencol$ in the velocity range of 7.9--13.9 $\kms$ superposed on the $\thirteencol$ integrated intensity contour as shown in the Figure \ref{integ_int}b. The lowest contour and contour intervals are the same as Figure \ref{integ_int}b. The black and white cross represent the positions of HD~37903 and NGC~2023 MM1, respectively. Black dots a--c in each map show the locations of the CO spectra in Figure \ref{spec}.}
  \label{mom}
\end{figure*}

\begin{figure*}
  \begin{center}
    \includegraphics[width=\linewidth]{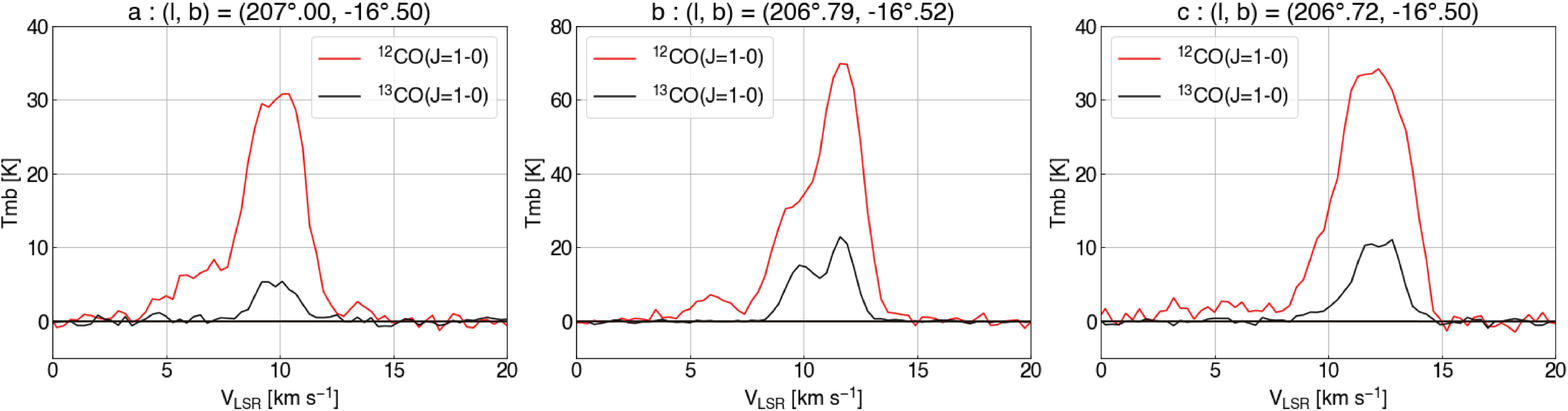}
  \end{center}
 \caption{CO spectra at the positions of the dots a--c in Figure \ref{mom}. $\twelvecol$ and $\thirteencol$ are plotted in black and red, respectively.}
   \label{spec}
\end{figure*}

Figure \ref{mom} shows distributions of the 1st moment and 2nd moment of the $\thirteencol$ emission. We see a clear trend that the cloud is blue-shifted in the east and is red-shifted in the west in addition to another red-shifted component at $(l$, $b)$ = $(207\fdg0$, $-16\fdg7$). The red-shifted gas shows an arc like distribution at $l$ = 206.65 to 206.75 degree, and $b = -16.6$ to $-16.4$ degree. Within the intensity level above 30 $\kkms$, the 2nd moment is enhanced toward the region where the 1st moment is 11 $\kms$ at $(l$, $b)$ = ($206\fdg8$, $-16\fdg5$), and another enhancement, while less obvious, is seen at $(l$, $b)$ = ($206\fdg93$, $-16\fdg6$). Figure \ref{spec} shows three CO profiles at the positions a, b, and c denoted in Figure \ref{mom}, where we find the velocity peaks consistent with Figure \ref{mom}. The profiles are peaked at 10 $\kms$ at position a, and at 12 $\kms$ at position c. The $\thirteencol$ profile at position b shows two peaks at 10 $\kms$ and 12 $\kms$, which are not clear in $\twelvecol$. This indicates that $\twelvecol$ is partially saturated due to high optical depth. The 5 $\kms$ cloud is unrelated with the present cloud and is not discussed further (\cite{2019arXiv191211607E}).

The above results show that there are two velocity components in the NGC~2023 region. \citet{2019arXiv191211607E} pointed out that the two clouds are mostly merged in NGC~2023, while they are resolved into two at 10 $\kms$ and 13 $\kms$ at some places. This is consistent with the present results. We adopt the velocity ranges based on the moment method used by \citet{2019arXiv191211607E} for the the blue-shifted cloud (the blue cloud) and red-shifted cloud (the red cloud). The 1st moment map in Figure \ref{mom} was used to separate the area with less than 10 $\kms$ and that more than 12 $\kms$, and the average values of the first moment and the second moment are calculated. The mean velocity and velocity dispersion are then calculated. The upper bound of the blue cloud and the lower bound of the red cloud are about 10.85 $\kms$, corresponding to the dip between the two peaks in the $\thirteencol$ profile in Figure \ref{spec}b. Further, in Figure \ref{chmap} the cloud distribution changes significantly at around the velocity. We define 10.85 $\kms$ as the boundary between the blue cloud and the red cloud.

\subsection{Complementary distribution between the blue-shifted and red-shifted clouds}

\begin{figure}
  \begin{center}
    \includegraphics[width=8cm]{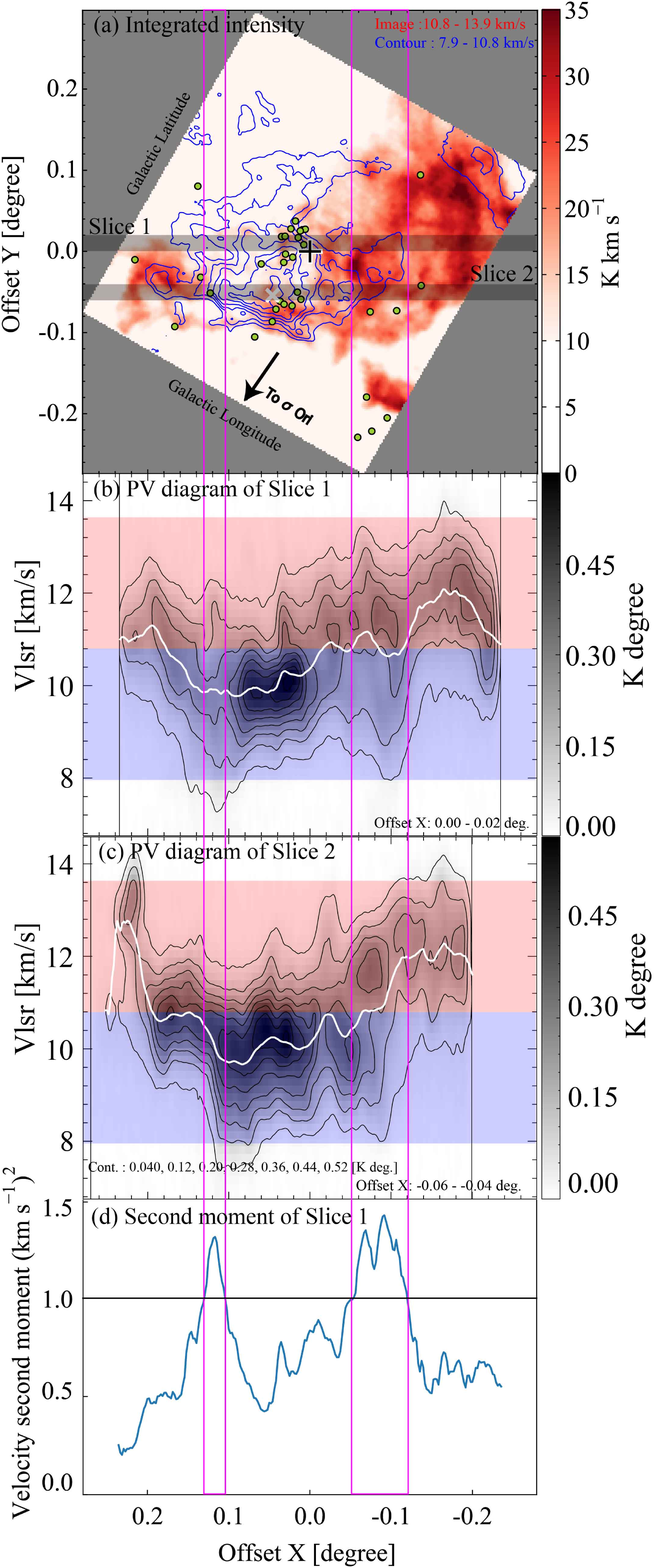}
  \end{center}
  \caption{(a) The $^{13}$CO($J$ = 1--0) distribution of red and blue clouds in the offset X--Y coordinates. The offset X--Y coordinate is defined by rotationg the galactic coordinate clockwise by 30.00 degree. The rotation center is the position of HD~37903. The image shows the $\thirteencol$ intensity integrated over the velocity range from 10.8--13.9 $\kms$. The blue contours show the intensity integrated over the range of 7.9--10.8 $\kms$. (b) The offset X--velocity diagram of the NGC~2023 region in $\thirteencol$. Integration range in the offset Y is indicated as dark transparent belt superposed on the panel (a). The overlaid white line indicates velocity first moment in the same region as the offset X--velocity diagram averaged along the offset Y. (c) The graph presents the second moment in the same region as (b) averaged along the offset-Y. The vertical purple dotted line shows the region where the second moment in the (c) is larger than 1.0 $(\kms)^2$.}
  \label{complementary}
\end{figure}

Figure \ref{complementary}a shows an overlay of the two velocity components at 10 $\kms$ and 12 $\kms$, where the two components show complementary distribution. Figures \ref{complementary}b and \ref{complementary}c show position-velocity diagrams along the two straight lines in Figure \ref{complementary}a. The lines were adopted so as to include the two red-shifted components on the east and west as well as the position close to the stellar cluster including the B star. In the two diagrams we find a velocity distribution with a V-shape. The velocity variation which is consistent with that in Figure \ref{complementary}a is seen. We find a hint that the linewidths are enhanced toward the two positions between the two velocity components at offset $X = \pm 0.1$ degrees, which are consistent with the enhanced 2nd moment in Figure \ref{complementary}d. Table \ref{table:extramath} lists the physical parameters of the two clouds, average column density $N(\mathrm{H_2})$, the cloud mass $M$ and the velocity range. The column density $N(\mathrm{H_2})$ is calculated by using the CO-to-H$_2$ conversion factor ($X_\mathrm{CO}$) as follows;
\begin{equation}
\nhtwo = W(\mathrm{CO})X_{\mathrm{CO}},
\label{xco}
\end{equation}
where $W(\mathrm{CO})$ ($\kkms$) is the integrated intensity of the CO emission. The molecular mass is given by equation (\ref{nh}), 
\begin{equation}
M = m_{\mathrm{H}} \mu D^2\Omega \sum_{i}N_i(\mathrm{H_2}),
\label{nh}
\end{equation}
where $m_{\mathrm{H}}$ is the mass of hydrogen molecule, $\mu$ is the mean molecular weight, $D$ distance to NGC2023 410 pc (\cite{2007A&A...474..515M}), $\Omega$ the solid angle of a pixel, and $N(\mathrm{H_2})$ the column density in eq. (\ref{xco}). The CO-to-H$_2$ conversion factor $X_{\mathrm{CO}}$ is taken as $1.0 \times 10^{20}$ (\kkms)$^{-1}$ cm$^{-2}$ (\cite{2017ApJ...838..132O}), and only the pixels with $W(\mathrm{CO})$ larger than 5$\sigma$ are used. Although the $\twelvecol$ emission is self-absorbed locally, the positions with clear self-absorption are limited in area. We thus adopted the method.

\begin{table*}[h]
\tbl{Column densities and masses of total cloud, blue cloud and red cloud.}{%
\begin{tabular}{lccccccccc}
\hline\noalign{\vskip3pt} 
\multicolumn{1}{c}{Cloud Name} & $V_\mathrm{LSR}$ & Column Density & Mass \\
 & (\kms) & (cm$^{-2}$) & (\msun) &  \\  [2pt] 
\multicolumn{1}{c}{(1)} & (2) & (3) & (4) &  \\  [2pt] 
\hline
\hline\noalign{\vskip3pt} 
Blue cloud & 7.9 --10.8& $1.6\times10^{22}$ &  $6.0\times10^2$ &\\
Red cloud & 10.8 --13.9 & $1.3\times10^{22} $&  $5.0\times10^2$ & \\
Total & 7.9 --13.9 & $2.6\times10^{22}$ &  $1.1\times10^3$ & \\
\hline\noalign{\vskip3pt} 
\end{tabular}}\label{table:extramath}
\begin{tabnote}
Note. --- Col. (1): Cloud name. Cols. (2) The velocity range of each cloud in Local Standard of Rest coordinate. Cols. (3) The maximum molecular hydrogen column density $N$(H$_2$) derived from the $^{12}$CO($J$ = 1--0) integrated intensity and the CO-to-H$_2$ conversion factor ($X_{\mathrm{CO}}$) of $1.0\times10^{20}$ (\kkms)$^{-1}$ cm$^{-2}$ (\cite{2017ApJ...838..132O}) Col. (4): Mass of the clouds derived using the equation (\ref{nh}). We used the integrated intensity higher than 5$\sigma$ in calculation. 
\end{tabnote}
\end{table*}

\begin{figure*}[]
  \begin{center}
    \includegraphics[width=18cm]{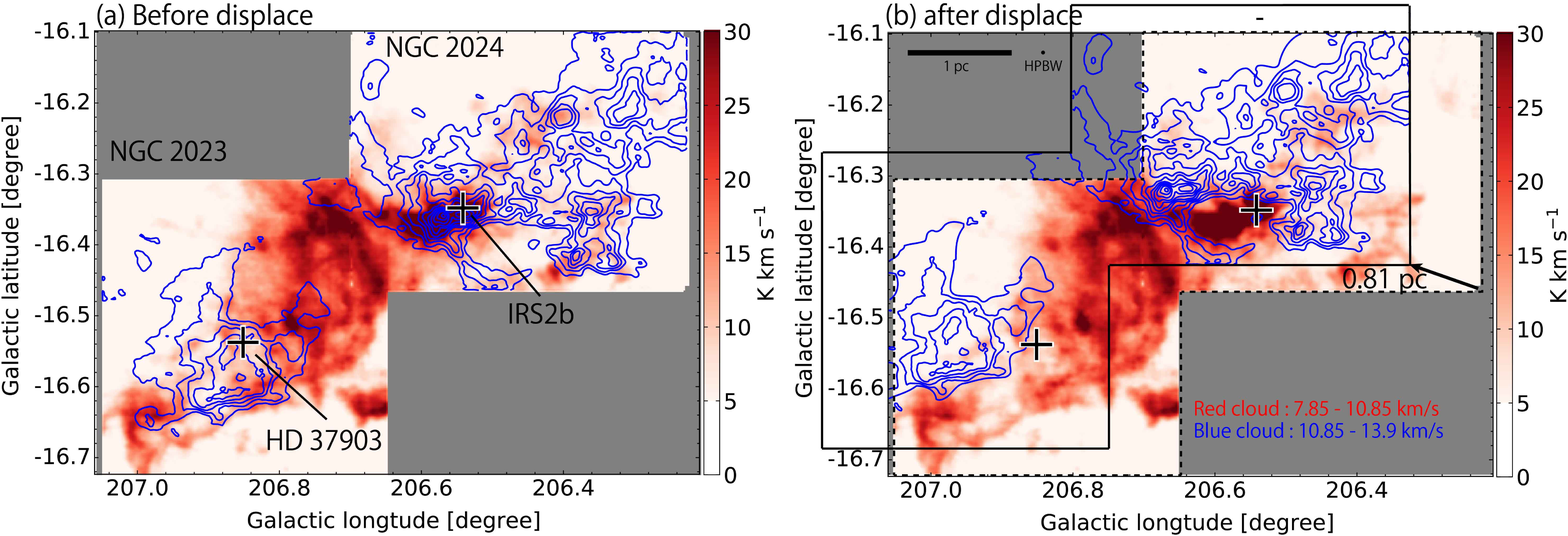}
  \end{center}
  \caption{(a) Integrated intensity covering NGC~2023 and NGC~2024 region in $\thirteencol$. Image shows the red shifted component of 10.85--13.9 $\kms$. Overlaid con
  contour represent the distribution of blue-shifted component of 7.9--10.85 $\kms$. (b) Integrated intensity distribution of red and blue components displaced for 0.81 pc.}
  \label{ponchi}
\end{figure*}

\section{Discussion}\label{sec:discussion}
\subsection{A cloud-cloud collision scenario}
Based on the results in Section \ref{sec:results} we frame a hypothesis that the NGC~2023 cloud consist of two clouds of projected velocity difference of 2 $\kms$. The clouds are likely colliding to trigger the formation of the B1.5 stars as well as the twenty YSOs in the two clusters. In the following, we describe details of the collisional process inferred from the present data and compares the results with the cloud-cloud collisions in the other regions, in particular, with the NGC~2024 cloud which seems to be connected to the present clouds (\cite{2008hsf1.book..662M}). 

The two regions, NGC~2023 and NGC~2024 are connected as part of a large cloud extending to the north and the two velocity components are common properties in the NGC~2023 and NGC~2024 regions (for the whole view of the CO distribution see \cite{2015ApJS..216...18N}). Figure \ref{ponchi} shows a combined $\thirteencol$ image of the NGC~2023 and NGC~2024 regions for the two velocity ranges of the colliding clouds taken with the 45 m telescope. A full account of the NGC~2024 region will be presented elsewhere (Enokiya et al. in preparation). \citet{2019arXiv191211607E} showed that the two clouds in NGC~2024 have complementary distribution with each other with a displacement of 0.6 pc based on the $\thirteencoh$ data taken with NANTEN2. We applied a displacement for the two regions in the same direction by assuming that the two clouds share common relative motion moving to east along the Galactic plane in a large scale in Figure \ref{ponchi}b. Figure \ref{ponchi}b shows the complementary distribution of 0.81 pc seems fairly good. The direction of the displacement explains that the site of star formation is on the western side of blue cloud in NGC~2023. The displacement is somewhat larger than in \citet{2019arXiv191211607E} probably due to the higher resolution. So, the picture is that the whole parent clouds, both red-shifted and blue-shifted, extended by $\sim$10 pc are commonly triggering high mass star formation in the two interaction sites toward NGC~2024 and NGC~2023. We shall adopt the picture in the present paper, while the weakness of the CO emission in the northeast makes the complementarity in NGC~2023 somewhat obscure as compared with NGC~2024. The collision time scale is roughly estimated to be 0.4 Myr from a ratio of 0.8 pc/2 $\kms$ for an assumed angle of the collision direction to the line of sight of 45 degrees.

Star formation is taking place in the red cloud in NGC~2024 and in the blue cloud in NGC~2023. This suggests that density was higher in each cloud in the regions (\cite{2015MNRAS.450...10H}). A marked difference is the number of high-mass stars between the two regions; more than ten high-mass stars in NGC~2024 and only one early B star in NGC~2023. This suggest that the column density in NGC~2024 was ten times higher in NGC~2024 than in NGC~2023 (see Figure 9 of \cite{2019arXiv191211607E})

Enhancement of the 2nd moment is seen toward the two parts between the two clouds in Figure \ref{complementary}d. This is understood as caused by the enhanced turbulence in the collisional interface layers in the east and west of the blue-shifted cloud. The collision is able to compress gas according to the process as simulated by theoretical works (\cite{1992PASJ...44..203H}; \cite{2013ApJ...774L..31I}; \cite{2014ApJ...792...63T}; \cite{2018PASJ...70S..54S}). It is probable that the \thirteencol~core, having a size of ~0.5 pc $\times$ 0.8 pc was formed by the compression. The typical column density is $\sim$$7 \times 10^{21}$ cm$^{-2}$ and the compressed mass is $\sim$100 \msun, where B1.5 star and 20 YSOs were formed. The total stellar mass of these stars is estimated to be $\sim$30 \msun (= 10 \msun $+$ $20\times1$ \msun), corresponding to a star formation efficiency of 30$\%$. The YSOs may be still growing by mass accretion in the $\thirteencol$ core, while the B1.5 star has dispersed the ambient gas by its winds and created the cavity with the reflection nebula as discussed below. The elongation of the YSOs and the $\thirteencol$ core nearly vertical to the collision direction is consistent with collision nearly along the Galactic plane, which is similarly to the collision in NGC~2024 (\cite{2019arXiv191211607E}). We presume that the collision path of the blue-shifted cloud is somewhat tilted to the west, resulting stronger compression on the western edge of the blue shifted cloud. This also explains the arc shape of the red-shifted cloud as caused by the bow-shock by the collision.   

\subsection{The cavity toward HD~37903}
\begin{figure*}
  \begin{center}
    \includegraphics[width=\linewidth]{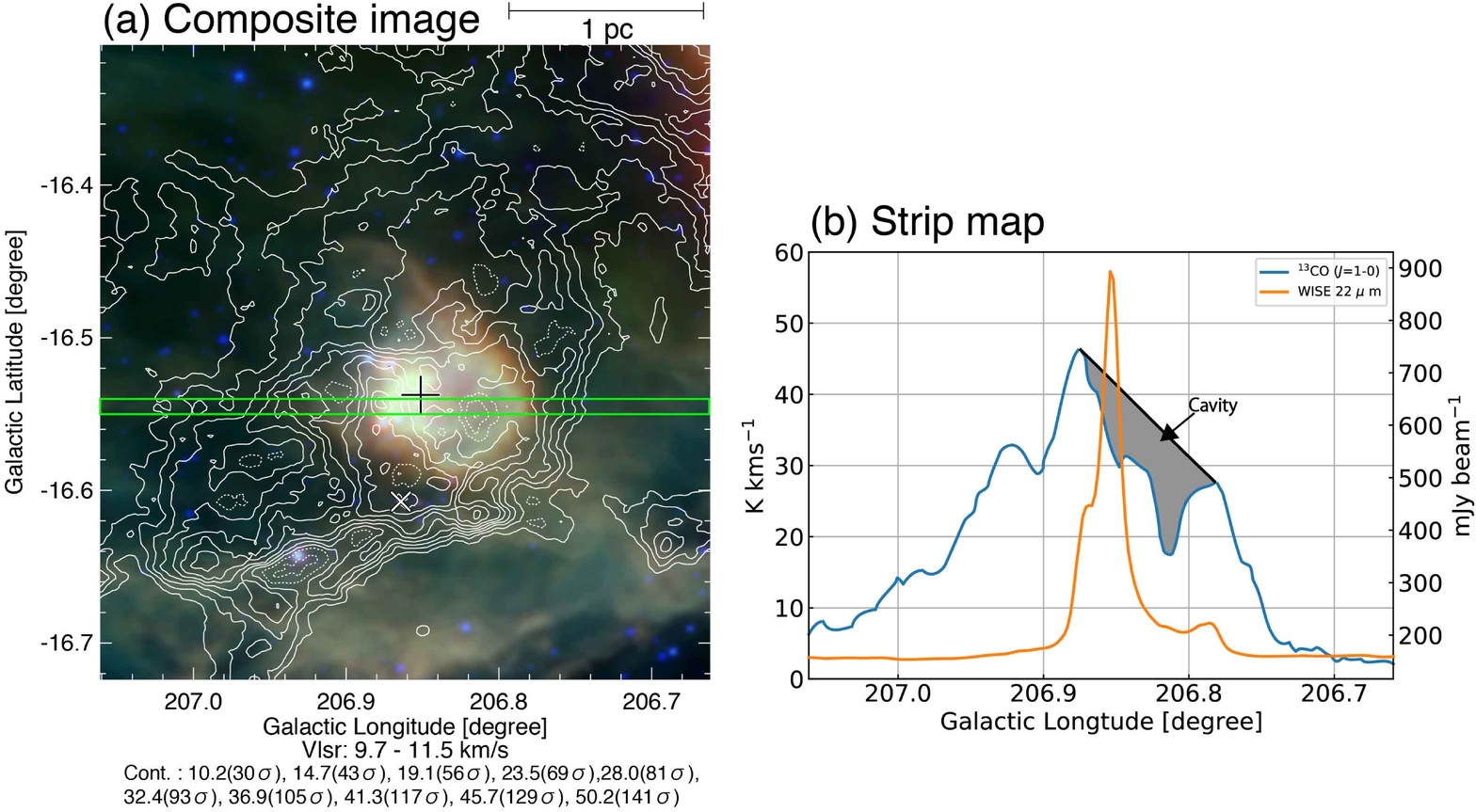}
  \end{center}
  \caption{(a) Three-color infrared image of NGC~2023 obtabined with WISE. The red, green, and blue represent 22 $\mu$m, 12 $\mu$m, and 4.6 $\mu$m, respectively. Contours show total intensity of $\thirteencol$ with the velocity between 7.9 and 13.9 $\kms$. The black cross and white cross represent the positions of HD~37903 and NGC~2023 MM1, respectively. (b) Intensity of the $\thirteencol$ and 22 $\mu$m on the green line superposed on the panel (a).}
  \label{ir}
\end{figure*}

Figure \ref{ir}a shows an overlay of the $\thirteencol$ core with the infrared image taken with WISE, and Figure \ref{ir}b shows a strip map of the infrared emission and the \thirteencol ~intensity along the green line in Figure \ref{ir}a. The infrared emission is the reflection light of the B1.5 star. The reflected light is asymmetric with respect to the star, which is probably due to the $\thirteencol$ gas distribution just after the collisional compression which had a peak toward $l = 206\fdg87$ with density decrease toward the west. We speculate that the distribution caused the present asymmetric cavity with respect to the B1.5 star, and that the stellar winds pushed the gas more deeply into the western part with density decrease. It is probable that the initial density distribution is more enhanced toward the current cavity and part of the gas was removed by the winds at $l = 206\fdg78$--$206\fdg85$. If we assume the initial density of $3.0\times 10^3$ cm$^{-3}$ with a size of 0.7 pc the total mass dispersed is estimated to be 24 $\msun$. So, currently the stellar feedback has a limited effect on the parent gas of 1000 $\msun$ at a level of a few \% in gas removal.

\subsection{The effect of the H{\sc ii} region in star formation}
\begin{figure*}
  \begin{center}
    \includegraphics[width=\linewidth]{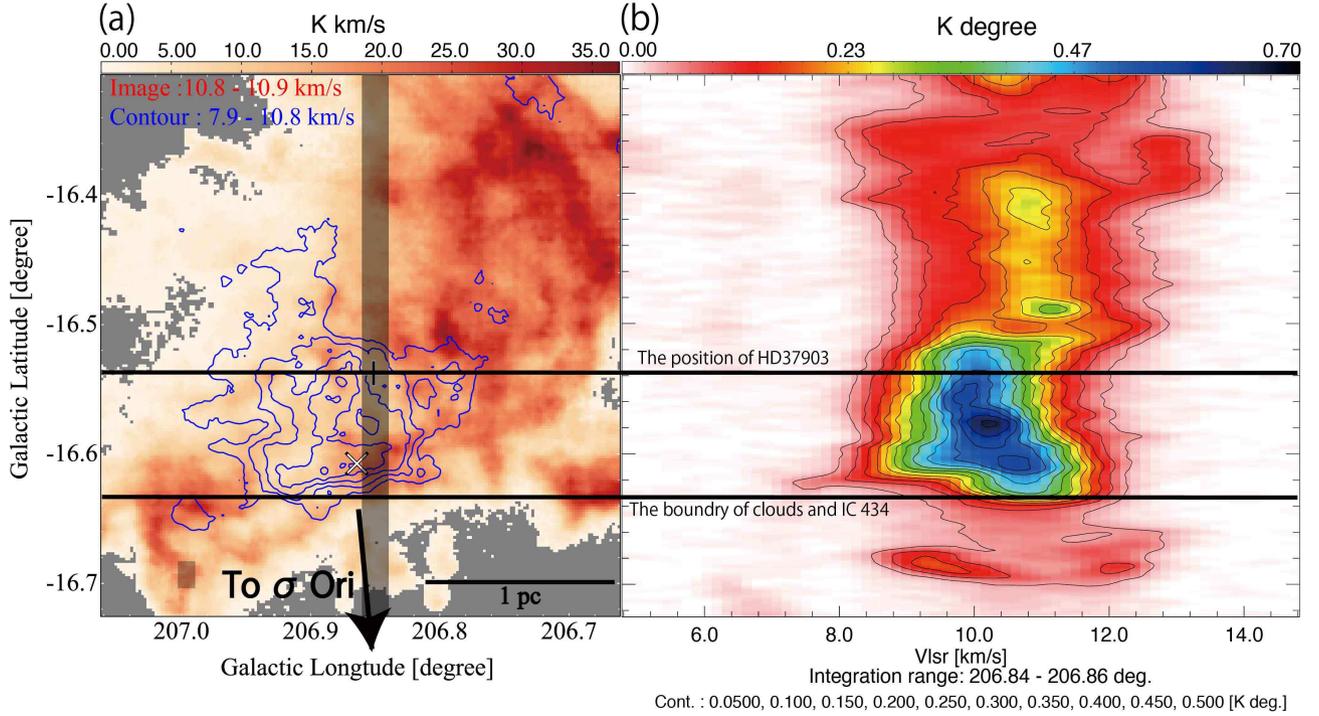}
  \end{center}
  \caption{(a) The $^{13}$CO($J$~=~1--0) distribution of red and blue clouds. The image shows the $\thirteencol$ intensity integrated over the velocity range from 10.8--13.9 $\kms$. The blue contours show the intensity integrated over the range of 7.9--10.8 $\kms$. The dark transparent belt shows the integration range of the Galactic longtude--velocity diagram presented in (b) The black arrow indicates the direction of the $\sigma$ Ori, which is the exciting star of the H{\sc ii} region IC~434. The upper and lower black vertical lines show the Galactic latitude of the exciting star HD~37903 and the boundary of molecular gas and the H{\sc ii} region.}
    \label{hii}
\end{figure*}

It has been a concern if IC~434 in triggering star formation in NGC~2023 and NGC~2024 (e.g., \cite{2009A&A...507.1485M}). Figure \ref{hii} shows a latitude-velocity diagram taken along the line at $l = 206\fdg85$ in Figure \ref{hii}. This diagram shows the velocity distribution of the $\thirteencol$ gas vertical to the ionization front formed by the H{\sc ii} region. The previous paper suggested such interaction may trigger star formation in NGC~2023 (\cite{2009A&A...507.1485M}). 

We find a strong intensity gradient toward $\sigma$ Ori as is consistent with the effect of the ionization. In this region the boundary between IC~434 and the CO gas is seen as the optical emission nebula of H{\sc ii} gas, while the exact location of $\sigma$ Ori is not known. It is possible that there is some velocity shift in the CO gas facing $\sigma$ Ori if the gas is accelerated by the H{\sc ii} region. It is however not clear how the gas is accelerated by the ionization front, because no significant velocity shift is found at Galactic latitude above $-16\fdg55$ in Figure \ref{hii}. We see some hint of a velocity shift only at latitude below 16.6 degrees, where a small shift from 10 $\kms$ to 11 $\kms$ may exit in the $\thirteencol$ cloud. Most of the star forming $\thirteencol$ cloud, therefore, seems to be unaffected. It is also possible the small shift is due to the collisional acceleration with the red-shifted cloud in the south, where the remnant of the red-shifted cloud has been ionized significantly below $-16\fdg63$. Consequently, we find no compelling evidence for the trigger of star formation by the H{\sc ii} region.

\section{Conclusions}\label{sec:conclusions}

The region of NGC 2023 was studied by using the $\twelvecol$ and $\thirteencol$ data taken with the NRO 45 m telescope at 19$\arcsec$ resolution. The data allowed us to reveal detailed gas kinematics at high spatial/velocity resolution of 0.04 pc / 0.3 km s$^{-1}$ at a distance of 410 pc. We have successfully resolved the interacting two clouds and presented a scenario of CCC as a trigger of star formation in NGC~2023. The main conclusions are summarized below.
\begin{enumerate}

\item The cloud consists of two velocity components at 10 $\kms$ and 12 $\kms$, and the two clouds show complementary distribution. The masses of the blue-shifted cloud and the red-shifted cloud are estimated to be 600 \msun and 500 \msun, respectively, with a typical column density of (5--6$) \times 10^{21}~$cm$^{-2}$. We find broadening of the linewidth toward the two interfaces of the two clouds which is a possible signature of the dynamical interaction.

\item We present a hypothesis that the two clouds collided with each other and triggered the formation of a B1.5 star HD~37903 and twenty young stellar objects. The $\thirteencol$ core has $\sim$100 $\msun$ and the star formation efficiency is estimated to be $\sim$30\%. The gas column density of the core $5\times10^{21}$ cm$^{-2}$ is slightly below the empirical threshold for O star formation, and the formation of a B1.5 star is consistent with the threshold. This shows that cloud-cloud collision also triggers not only high mass stars but also the formation of low-mass stars. The time scale of the collision is estimated to be 0.2~Myr. This short time is consistent with the stellar age as well as that there is no displacement in the complementary distribution.

\item The NGC~2023 cloud having a size of roughly 2~pc is elongated nearly along the Galactic plane and is physically connected with the NGC~2024 cloud which is separated by 4 pc from each other. A comparison with the cloud-cloud collision in NGC~2024 (\cite{2019arXiv191211607E}) suggests that the collisions are taking in a similar manner with each other in terms of the collision direction nearly parallel to the plane. 
\end{enumerate}

Including the other regions in Orion where high star formation was triggered by cloud-cloud collision, M42/M43 (\cite{2018ApJ...859..166F}), NGC~2071/NGC~2069 (\cite{2020PASJ..tmp..163F}), and NGC 2024 (\cite{2019arXiv191211607E}), we infer that cloud-cloud collision triggered all the high-mass star formation in Orion A and B. We suggest that cloud-cloud collision provides a viable alternative to the sequential star formation scenario (\cite{1977ApJ...214..725E}).

\begin{ack}
We are grateful to Akio Taniguchi, Kazuki Shiotani, Keisuke Sakasai, and Kenta Matsunaga for their valuable support during data analysis. We also acknowledge Hiroaki Yamamoto for useful discussion of this paper. The Nobeyama 45-m radio telescope is operated by Nobeyama Radio Observatory, a branch of National Astronomical Observatory of Japan. This publication makes use of data products from the Wide-field Infrared Survey Explorer, which is a joint project of the University of California, Los Angeles, and the Jet Propulsion Laboratory/California Institute of Technology, funded by the National Aeronautics and Space Administration. This study was financially supported by Grants-in-Aid for Scientific Research (KAKENHI) of the Japanese Society for the Promotion of Science (JSPS; grants No., 15H05694, 25287035, 19H05075, and 20H01945). H.S. was supported by ``Building of Consortia for the Development of Human Resources in Science and Technology'' of Ministry of Education, Culture, Sports, Science and Technology (MEXT; grant No. 01-M1-0305). 
\end{ack}

\end{document}